\documentclass[12pt]{iopart}

\usepackage{bm}
\expandafter\let\csname equation*\endcsname\relax
\expandafter\let\csname endequation*\endcsname\relax
\usepackage{amsmath,amssymb}
\usepackage{CJK}
\usepackage{hyperref}
\usepackage{graphicx}
\usepackage{cite}
\begin{document}
\begin{CJK*}{GBK}{ }
\title{Weak and Strong Coupling Polarons in Binary Bose-Einstein Condensates}

\author{Ning Liu }

\address{School of Mathematics and Physics, Anqing Normal University, Anqing 246133, China}
\ead{ningliu@mail.bnu.edu.cn}
\vspace{10pt}
\begin{indented}
\item[]
\end{indented}

\begin{abstract}
The Bose polaron is a quasiparticle that arises from the interaction between impurities and Bogoliubov excitation in Bose-Einstein condensates, analogous to the polaron formed by electrons and phonons in solid-state physics. In this paper, we investigate the effect of phase separation on weakly coupled and strongly coupled Bose polarons. Our findings reveal that phase separation induces a remarkable alteration in the properties of weakly coupled Bose polarons. However, in the case of strong coupling, phase separation cannot destroy the highly self-trapping state of Bose polarons.
\end{abstract}

%
%
%
%
%

\section{Introduction}

Polarons are foundational concepts in quantum many-body theory, representing quasiparticles arising from electron-phonon interactions~\cite{Mahan2000,Devreese2009,Chatterjee2019}. However,  an interesting question about polarons has not been solved, i.e., the existence of polarons near quantum critical points. In recent years, the concept of polarons has been greatly expanded, particularly with the proposal of Bose polarons~\cite{Grusdt2015,Huang2009, Novikov2009, Tempere2009, Rath2013, Ardila2015, Shchadilova2016,Volosniev2017,Guenther2018, Ardila2019, Ichmoukhamedov2019,Drescher2019, Field2020, Isaule2021, Khan2021, Boud2014,Boud2015,Camacho2023}.  Bose polarons are quasiparticles that emerge when an impurity modifies a host atom within a Bose-Einstein condensate (BEC). Notably, Bose polarons have been observed experimentally~\cite{Hu2016,Jorgensen2016,Skou2021,Skou2022} and  provide an excellent platform to investigate whether quasiparticles exist near quantum critical points~\cite{Yan2020}.

With advancements in experimental techniques for ultracold atomic gases, it has become feasible to create multi-component quantum mixtures~\cite{Lamporesi2023}, opening up new avenues for studying polaron behavior at phase boundaries. By manipulating the magnetic field, interactions in a two-component Bose-Einstein condensate can be modified, facilitating the transition from a miscible phase to an immiscible phase~\cite{Pitaevskii2016,Boud2020}. This presents an opportunity to investigate the behavior of Bose polarons near the transition. This inspires us to explore the effect of phase separation on the properties of Bose polarons.

In the previous study, the authors employed the Lee-Low-Pines(LLP) theory to explore the properties of weakly coupled polarons in two-component Bose-Einstein condensates~\cite{Liu2023}. In the current study, we adopt a modified approach called the Lee-Low-Pines-Huybrechts (LLPH) method ~\cite{Huybrechts1977,Mukhopadhyay1999,Kervan2003} to overcome the limitations of the LLP theory when dealing with strongly coupled scenarios. The LLPH method combines the LLP theory and the Landau-Pekar theory. In particular, the comparison between the LLPH method and the Feynman path integral method shows its good accuracy~\cite{Huybrechts1977}. By conducting the calculation, we confirm that the obtained results are in agreement with the previous result in the weak coupling regime. This implies that the properties of weakly coupled Bose polarons can undergo significant changes due to phase separation. Whereas, in strongly coupled cases, we discover that the self-trapping state of Bose polarons is hardly affected by phase separation.

The paper is organized as follows: In Sec.\ref{s1}, we derive the ground state of polarons in a uniform single-component dilute Bose gas. In Sec.\ref{s2}, we show the effect of phase separation on the Bose polaron. Finally, Sec. \ref{con} provides a summary and conclusion.

\section{Polarons in a single-component BEC}\label{s1}
In this section, we derive the Bose polaron model and show the ground state energy results by the LLPH method. The energy expression obtained is consistent with the results obtained in the previous literature.

We consider the model of impurity coupling with a uniform single bosonic bath, which is characterized by the general Hamiltonian,
\begin{equation}
  H=\frac{p^2}{2m_I}+\sum_{\bm{k}}\epsilon_ka^\dagger_{\bm{k}}a_{\bm{k}}+\frac{1}{2\Omega}\sum_{\bm{pqk}}g_{BB}a_{\bm{p}+\bm{k}}^\dagger a^\dagger_{\bm{q}-\bm{k}}a_{\bm{q}}a_{\bm{p}}+\frac{1}{\sqrt{\Omega}}\sum_{\bm{pk}}g_{BI}\rho(\bm{k})a^\dagger_{\bm{p}-\bm{k}}a_{\bm{p}}.\label{gh}
\end{equation}
Here, $m_I$ and $p_I$ denote the mass and momentum of the impurity atom, $\epsilon_k$ represents the kinetic energy of host atoms,  $\epsilon_k=\hbar^2k^2/2m$, where $m$ is the mass of the host atom. $\Omega$ represents the volume of the condensate. $a^\dagger_{\bm{k}}$($a_{\bm{k}}$) represents the creation(annihilation) operators of the host boson. $g_{BI}$ characterizes the contact interactions between impurities and Bose atoms, and $g_{BB}$ represents the contact interaction of Bose atoms. Specifically, $g_{BI}={2\pi\hbar^2a_{BI}}/{\mu}$ and $g_{BB}={4\pi\hbar^2a_{BB}}/{m}$, where $a_{BI}$ and $a_{BB}$ are the scattering lengths for impurity-boson and boson-boson interactions, respectively. In this article, we only consider positive scattering lengths for all interactions. The induced mass is defined as $\mu={mm_I}/{(m+m_I)}$. The impurity density is given by
\begin{equation}
  \rho(\bm{k})=\int d\bm{r}'e^{{\rm i}\bm{k}\cdot\bm{r}}\delta(\bm{r}-\bm{r}').
\end{equation}

 At $T=0$, the Bose gas condenses, which means that it has a macroscopic occupation in the lowest energy state. The phenomenon introduces the Bogoliubov approximation. The diagonalized Hamiltonian is derived,
 \begin{equation}
\begin{aligned}
  H_d&=\frac{p^2}{2m_I}+g_{BI}n_0+\sum_{\bm{k}}\varepsilon_k\alpha^\dagger_{\bm{k}}\alpha_{\bm{k}}+\sum_{\bm{k}}\sqrt{n_0}V_{BI}(\bm{k}){\rm e}^{{\rm i}\bm{k}\cdot \bm{r}}(\alpha_{\bm{k}}+\alpha^\dagger_{-\bm{k}})\\
  &+\mbox{extended Fr\"{o}hlich terms},\label{efh}
  \end{aligned}
\end{equation}
through the Bogoliubov approximation, i.e., $a_{\bm{0}}\rightarrow \sqrt{N_{0}}$, and the Bogoliubov transformation~\cite{Huang2009,Tempere2009},
\begin{equation}
  \alpha_{\bm{k}}=u_ka_{\bm{k}}+v_{k}a^\dagger_{-\bm{k}},
\end{equation}
where $u_k,v_k=\pm\sqrt{(\epsilon_{k}+g_{BB}n)/2\varepsilon_k\pm1/2}$.
The density of the condensed atoms is given by $n_0=N_0/\Omega$. $\alpha_{\bm{k}}(\alpha^\dagger_{\bm{k}})$ represents the annihilation (creation) operator for the Bogoliubov excitation. The Bogoliubov excitation spectrum and effective interaction are given by:
\begin{equation}
  \varepsilon_k=\sqrt{\epsilon_k^2+2g_{BB}n_0\epsilon_k},\  V_{BI}(\bm{k})= g_{BI}\sqrt{\frac{\epsilon_k}{\varepsilon_k}}.
\end{equation}
The fourth term in Eq.(\ref{efh}) represents the coupling of a single phonon excitation with an impurity, while the extended coupling term indicates the presence of multiple phonon excitation.

The LLPH method proposed by Huybrechts combines Landau-Pekar's strong coupling theory and LLP theory~\cite{Huybrechts1977}. First, the ordinary LLP transformation is generalized, $H'=U^{-1}_1H_dU_1$, where
\begin{equation}
  U_1={\rm exp}\left[{-{\rm i}\sigma\sum_{\bm{k}}\bm{k}\cdot \bm{r}\alpha^\dagger_{\bm{k}}\alpha_{\bm{k}}}\right],\label{at}
\end{equation}
 Here, $\sigma$ is a parameter. And then the coordinates and momenta of the impurity switch to the description of the creation and annihilation operators with a variational parameter,
\begin{equation}
  \begin{aligned}
    p_j&=\sqrt{\frac{\hbar m_I\lambda}{2}}\left(b_j^\dagger+b_j\right),\\
    r_j&={\rm i}\sqrt{\frac{\hbar}{2 m_I\lambda}}\left(b_j-b_j^\dagger\right).\label{cao}
  \end{aligned}
\end{equation}
Here, $\lambda$ is a variational parameter and the subscript $j$ indicates three components. The Hamiltonian $H'$ becomes
 \begin{equation}
  \begin{aligned}
    H'&=g_{BI}n_0+\frac{\hbar\lambda}{2}\left(\sum_jb_j^\dagger b_j+\frac{3}{2}\right)+\sum_{\bm{k}}\left(\sigma^2\epsilon_k+\varepsilon_k\right)\alpha_{\bm{k}}^\dagger\alpha_{\bm{k}}\\
    &+\sum_{\bm{k}}V_{IB}(\bm{k})(\alpha^\dagger_{\bm{k}}+\alpha_{\bm{k}}){\rm e}^{-\frac{(1-\sigma)^2\hbar}{4m\lambda}k^2}{\rm e}^{-(1-\sigma)\sqrt{\frac{\hbar}{2m\lambda}}\sum_jb_j^\dagger k_j}{\rm e}^{(1-\sigma)\sqrt{\frac{\hbar}{2m_I\lambda}}\sum_j b_j k_j}\\
    &+\frac{\hbar\lambda}{4}\sum_j\left(b_j^\dagger b_j^\dagger+b_jb_j\right)+\frac{\sigma^2}{2m_I}\sum_{\bm{k},\bm{k}'}\hbar^2\bm{k}\cdot\bm{k}'\alpha^\dagger_{\bm{k}}\alpha_{\bm{k}'}^\dagger\alpha_{\bm{k}}\alpha_{\bm{k}'}\\
    &-\sum_{\bm{k},j}\sqrt{\frac{\hbar\lambda}{2m_I}}\left(b_j^\dagger+b_j\right)\alpha_{\bm{k}}^\dagger\alpha_{\bm{k}}+\mbox{higher terms}\label{H'}.
  \end{aligned}
\end{equation}
 Notice that the second line in Eq.(\ref{H'}) contains a Gaussian potential for the interaction term. That is why it is possible to describe strong coupling in the Landau-Pekar sense.

We introduce the variational wave function, $|\psi\rangle=U_2|0\rangle$, where $|0\rangle$ represents the phonon vacuum state and $U_2$ is a displacement operator acting on $\alpha_{\bm{k}}$,
\begin{equation}
  U_2={\rm exp}\left[{\sum_{\bm{k}}(\alpha^\dagger_{\bm{k}}f(k)-\alpha_{\bm{k}}f^*(k))}\right],
\end{equation}
where $f(k)$ is a variational function.

Before proceeding with the calculation, we need to address the issue of $\sigma$ as a variational parameter. When $\sigma\rightarrow 1$ in Eq.(\ref{at}), $U_1$ corresponds to the ordinary Lee-Low-Pines transformation~\cite{Lee1953}. On the other hand, for $\sigma\rightarrow0$, the method is equivalent to the Landau-Pakar method~\cite{Chatterjee2019}. The former corresponds to the case of weakly coupled polarons, while the latter corresponds to the case of strongly coupled polarons. The adiabatic limit and the localized state limit are necessary to consider here, thus $\sigma$ cannot be considered as a variational parameter over all the range, only in the range $0<\sigma<1$~\cite{Mukhopadhyay1999}.

The energy functional is given by $E_0=\langle\psi|H'|\psi\rangle$. By considering the condition $b_j|0\rangle=\alpha_{k}|0\rangle=0$, we obtain the energy functional,
\begin{equation}
\begin{aligned}
  E_0&=g_{BI}n_0+\frac{3}{4}\hbar\lambda\left(1-\sigma\right)^2+\sum_{\bm{k}}\left(\varepsilon_k+\sigma^2\epsilon_{Ik}\right)|f(k)|^2\\
  &+\sum_{\bm{k}}\sqrt{n_0}V_{BI}(\bm{k}){\rm e}^{-\frac{(1-\sigma)^2\hbar}{4m_I\lambda}k^2}\left[f(k)+f^*(k)\right].\label{E}
  \end{aligned}
\end{equation}
Minimizing the energy $E_0$ with respect to $f^*(k)$, we find
\begin{equation}
  f(k)=-\frac{\sqrt{n_0}V_{BI}(\bm{k}){\rm e}^{-\frac{(1-\sigma)^2\hbar}{4m_I\lambda}k^2}}{\varepsilon_k+\sigma^2\epsilon_{Ik}}.
\end{equation}
Substituting the expression for $f(k)$ into Eq.(\ref{E0}), the gound state energy is given by:
\begin{equation}
  E_0=g_{BI}n_0+\frac{3}{4}\hbar\lambda\left(1-\sigma\right)^2-\sum_{\bm{k}}\frac{n_0V^2_{BI}(\bm{k}){\rm e}^{-\frac{(1-\sigma)^2\hbar}{2m_I\lambda}k^2}}{\varepsilon_k+\sigma^2\epsilon_{Ik}}\label{E0}
\end{equation}
By converting the summation over $\bm{k}$ in Eq.(\ref{E0}) to an integral, we obtain
\begin{equation}
  \bar{E}_0=\left(1+\bar{m}\right)\frac{k_na_{BI}}{k_na_{BB}}+\frac{3\bar{m}}{2\bar{r}^2}\left(1-\sigma\right)^2-\gamma I_E\label{Ee}.
  \end{equation}
Here $\bar{m}={m}/{m_I}$, $\bar{r}=\sqrt{\hbar/m_I\lambda\xi^2}$, $\bar{E}_0=E_0/g_{BB}n_0$, $k_n=(6\pi^2n)^{1/3}$,
\begin{equation}
  \gamma=\frac{2}{\sqrt{3\pi^3}}\frac{(k_n a_{BI})^2}{(k_na_{BB})^{1/2}}(1+\bar{m})^2,
\end{equation}
and
\begin{equation}
 I_E= \int_0^\infty\frac{\bar{k}^2{\rm e}^{-\frac{(1-\sigma)^2\bar{r}^2}{2}\bar{k}^2}d\bar{k}}{\bar{k}^2+2+\sigma^2\bar{m}\bar{k}\sqrt{\bar{k}^2+2}}.
\end{equation}
In the above equation, $\bar{k}=k\xi$, where $\xi$ represents the healing length, $\xi=\sqrt{8\pi a_{BB}n}$. The variational parameter $\lambda$ is replaced by $\bar{r}$. In principle, we can obtain the equations satisfied by the variational parameters $\sigma$ and $\bar{r}$ by variations Eq.(\ref{Ee}). And then determine the values of the variational parameters by numerical calculation. But in certain limited cases of $\sigma$, it is possible to derive analytical expressions.

For the limit $\sigma\rightarrow 1$, Eq.(\ref{Ee}) is reduced to
\begin{equation}
   \bar{E}^{\rm weak}_0=\left(1+\bar{m}\right)\frac{k_na_{BI}}{k_na_{BB}}-\gamma \int_0^\infty\frac{\bar{k}^2d\bar{k}}{\bar{k}^2+2+\bar{m}\bar{k}\sqrt{\bar{k}^2+2}} \label{Ew}.
\end{equation}
We notice that the other variational parameter $\bar{r}$ in Eq.(\ref{Ee}) is eliminated in this case. It needs to be emphasized that the ground state energy becomes divergent in this limit. The divergence can be eliminated by introducing a higher-order term of the pseudopotential~\cite{Huang2009}. The numerical results of Eq.(\ref{Ee}) in the weak coupling regime are shown in ~\cite{Huang2009}. We express the result of Eq.(\ref{Ew}) as follows:
\begin{equation}
\bar{E}_0^{\rm weak}=\left(1+\bar{m}\right)\frac{k_na_{BI}}{k_na_{BB}}+\sqrt{2}\gamma\left[\frac{ \arccos\bar{m}}{\left(1-\left(\bar{m}\right)^2\right)^{3/2}}-\frac{\bar{m}}{1-\left(\bar{m}\right)^2}\right].
\end{equation}

In the limit of $\sigma\rightarrow 0$, Eq.(\ref{Ee}) describes the ground state energy of strongly coupled polarons. It is given by the following expression:
\begin{equation}
  \bar{E}^{\rm strong}_0=\left(1+\bar{m}\right)\frac{k_na_{BI}}{k_na_{BB}}+\frac{3\bar{m}}{2\bar{r}^2}-\gamma \int_0^\infty{\rm e}^{-\frac{\bar{r}^2}{2}\bar{k}^2}\frac{\bar{k}^2d\bar{k}}{\bar{k}^2+2}\label{Es}.
\end{equation}
The result of Eq.(\ref{Es}) is expressed as
\begin{equation}
  \bar{E}_0^{\rm strong}=\left(1+\bar{m}\right)\frac{k_na_{BI}}{k_na_{BB}}+\frac{3\bar{m}}{2\bar{r}^2}+\frac{\gamma}{\sqrt{2}}\left[{\pi}{\rm e}^{\bar{r}^2}{\rm erfc}(\bar{r})-\frac{\sqrt{\pi}}{\bar{r}}\right]\label{Ees}
\end{equation}
In Eq.(\ref{Ees}), ${\rm erfc}(\bar{r})$ represents the complementary error function defined as:
 \begin{equation}
  {\rm erfc}(\bar{r})=\frac{2}{\sqrt{\pi}}\int_{\bar{r}}^\infty{\rm e}^{-t^2}dt.
 \end{equation}
 For numerical results of Eq.(\ref{Ees}) in the strong coupling regime, we refer to the study~\cite{Casteels2011}.

\section{Polarons in two-component BECs}\label{s2}

In this section, we consider an impurity interacting with the homogeneous two-component BEC. When the phase separation condition is satisfied, that is, the intercomponent interaction is greater than the geometric average of intracomponent interactions, $|g_{+-}|>\sqrt{g_{++}g_{--}}$, the system will undergo a transition from miscible phase to immiscible phase. In order to ensure uniformity, we only consider the behaviors of the Bose polaron for a short time after the condition of phase separation is satisfied.

The Hamiltonian is derived from the general Bogoliubov transformation~\cite{Bighin2022,Liu2023}:
\begin{equation}
  \tilde{H}_F=\frac{p^2}{2m_I}+\sum_{\pm}g_{\pm I}n_{\pm0}+\sum_{\pm,\bm{k}}\varepsilon_{\pm}\alpha_{\pm,\bm{k}}^\dagger\alpha_{\pm,\bm{k}}+\sum_{\pm,\bm{k}}\sqrt{n_{\pm,0}}V_{\pm}e^{{\rm i }\bm{k}\cdot \bm{r}}\left(\alpha_{\pm, \bm{k}}+\alpha_{\pm,-\bm{k}}^\dagger\right)
\end{equation}
where,
\begin{equation}
  V_{\pm}=g_{\pm I}\sqrt{\frac{\epsilon_{\pm}}{\varepsilon_{\pm}}}\cos\theta_k\pm g_{\mp I}\sqrt{\frac{\epsilon_{\mp}}{\varepsilon_{\pm}}}\sin\theta_k.
\end{equation}
The plus and minus signs indicate the different components. The coupling $V_+$ is referred to as impurity-density (ID) coupling, while $V_-$ is impurity-spin (IS) coupling~\cite{Liu2023}. The angles $\theta_k$ are given by
\begin{align}
\cos\theta_k&=\sqrt{\frac{1}{2}\left[1+\frac{\varepsilon_+^2-\varepsilon_-^2}{\sqrt{(\varepsilon_+^2-\varepsilon_-^2)^2+16g_{+-}^2\epsilon_+\epsilon_-n_{+0}n_{-0}}}\right]},\\
\sin\theta_k&=\sqrt{\frac{1}{2}\left[1-\frac{\varepsilon_+^2-\varepsilon_-^2}{\sqrt{(\varepsilon_+^2-\varepsilon_-^2)^2+16g_{+-}^2\epsilon_+\epsilon_-n_{+0}n_{-0}}}\right]},
\end{align}
and the excitation spectrum $\varepsilon_{\pm}$ is expressed as:
\begin{equation}
  \varepsilon^2_{\pm}=\frac{1}{2}\left(\varepsilon_+^2+\varepsilon_-^2\right)\pm\frac{1}{2}\sqrt{\left(\varepsilon_+^2-\varepsilon_-^2\right)^2+16g_{+-}^2\epsilon_+\epsilon_-n_{+0}n_{-0}},
\end{equation}
where $\varepsilon_{\pm}=\sqrt{\epsilon_{\pm}^2+2g_{\pm}n_{\pm,0}\epsilon_{\pm}}$. For simplicity, we assume that the masses of the Bose atoms for two species are equal, so $\epsilon_{+}=\epsilon_-$. The couplings are defined as contact interactions,
\begin{equation}
  g_{\pm I}=\frac{2\pi\hbar^2 a_{\pm I}}{\mu},\ g_{\pm}=\frac{4\pi\hbar^2a_{\pm}}{m},\  g_{+-}=\frac{4\pi\hbar^2a_{+-}}{m},
\end{equation}
where $a_{\pm I}$, $a_{\pm}$, and $a_{+-}$ correspond to the impurity-component couplings, intra-component couplings, and inter-component couplings, respectively.

Similar to the case of the single-component BEC, we perform two transformations:
\begin{align}
  \tilde{U}_1&={\rm exp}\left[{-{\rm i}\sigma\sum_{\pm}\sum_{\bm{k}}\bm{k}\cdot \bm{r}\alpha^\dagger_{\pm,\bm{k}}\alpha_{\pm,\bm{k}}}\right],\\
   \tilde{U}_2&= {\rm exp}\left[{\sum_{\pm}\sum_{\bm{k}}(\alpha^\dagger_{\pm,\bm{k}}f_{\pm}(k)-\alpha_{\pm,\bm{k}}f^*_{\pm}(k))}\right] .\label{at2}
\end{align}
 The  energy functional is defined as $\tilde{E}_0={_-}\langle 0|{_+}\langle 0|\tilde{U}_2^{-1}\tilde{U}_1^{-1} \tilde{H}_F \tilde{U}_1\tilde{U}_2|0\rangle_+|0\rangle_-$, where $|0\rangle_{\pm}$ represent the phonon vacuum state for binary BECs. By minimizing the energy functional $\tilde{E}_0$, the variational parameter $f_{\pm}(k)$ can be determined:
\begin{equation}
  f_{\pm}(k)=-\frac{\sqrt{n_{\pm,0}}V_{\pm}(\bm{k}){\rm e}^{-\frac{(1-\sigma)^2\hbar}{4m_I\lambda}k^2}}{\varepsilon_{\pm,k}+\sigma^2\epsilon_{Ik}}.
\end{equation}
Therefore, the ground-state energy can be derived,
\begin{equation}
  \tilde{E}_0=\sum_{\pm}g_{\pm I}n_{\pm0}+\frac{3}{4}\hbar\lambda\left(1-\sigma\right)^2-\sum_{\pm,\bm{k}}\frac{n_{\pm,0}V^2_{\pm}(\bm{k}){\rm e}^{-\frac{(1-\sigma)^2\hbar}{2m_I\lambda}k^2}}{\varepsilon_{\pm}+\sigma^2\epsilon_{Ik}}\label{Egb}.
\end{equation}

In the limit of $\sigma\rightarrow 1$, the sum in Eq.(\ref{Egb}) can be evaluated to obtain the ground state energy in the weak coupling regime, which has been discussed in a previous work~\cite{Liu2023}. The results demonstrate the significant impact of phase separation on the properties of polarons. In particular, when phase separation occurs, the contribution from IS couplings to the ground energy decreases to zero, and the kinetic energy of the polaron tends to zero due to the divergent effective mass.

In the limit of $\sigma\rightarrow 0$, Eq.(\ref{Egb}) describes the ground energy of polarons in binary BECs under strong coupling:
\begin{equation}
  \tilde{E}_0^{\rm strong}=\sum_{\pm}\frac{k_na_{\pm I}}{k_na_{+}}+\frac{3\bar{m}}{2\bar{r}^2}-\gamma' \sum_{\pm} \left(\sqrt{1\pm\bar{a}}\pm\frac{a_{-I}}{a_{+I}}\sqrt{1\mp\bar{a}}\right)^2 I_{\tilde{E}},\label{bes}
\end{equation}
where
\begin{align}
  \gamma'&=\frac{2}{\sqrt{3\pi^3}}\frac{(k_n a_{+I})^2}{(k_na_{+})^{1/2}}(1+\bar{m})^2,\\
  \bar{a}&=\frac{a_+-a_-}{\sqrt{(a_{+}-a_-)^2+4a^2_{+-}}},\\
  \tilde{a}_{\pm}&=1+\frac{1}{a_{+}}\left[a_-\pm\sqrt{(a_{+}-a_{-})^2+4a_{+-}^2}\right],
\end{align}
and
\begin{equation}
  I_{\tilde{E}}=\int\frac{\bar{k}^2}{\bar{k}^2+\tilde{a}_{\pm}}{\rm e}^{-\frac{\bar{r}^2}{2}\bar{k}^2}d\bar{k}.
\end{equation}
It is worth noting that $\tilde{a}_->0$, i.e., $a_+a_->a_{+-}$, which implies the stability condition~\cite{Pitaevskii2016}. For $a_+a_-<a_{+-}$, the system undergoes a transition from a miscible phase to an immiscible phase, where phase separation occurs.  The result of $I_{\tilde{E}}$ is given by
\begin{equation}
   I_{\tilde{E}}=\frac{1}{2}\left[ \frac{\sqrt{2\pi}}{\bar{r}}-{\sqrt{\tilde{a}_{\pm}}\pi}{\rm e}^{\tilde{a}_{\pm}\bar{r}^2/2} {\rm erfc}\left(\sqrt{\frac{\tilde{a}_{\pm}}{2}}\bar{r}\right)\right].\label{IEe}
\end{equation}

 \begin{figure}[!htpb]
\centering
\includegraphics[width=10cm]{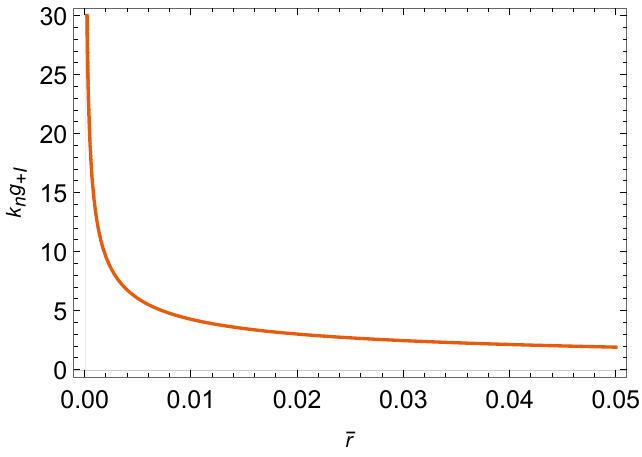}
\caption{(color online) Variational parameter $\bar{r}$ as a function of coupling strength $k_na_{+I}$. Here, $k_na_+=0.8$, $a_{+-}=0.9a_+$, $a_+=a_-$, $a_{+I}=a_{-I}$ and $\bar{m}=1/2$.}\label{ar}
\end{figure}
 Variation of Eq.(\ref{bes}) with respect to $\bar{r}$ yields the result shown in Figure \ref{ar}. We observe that in the case of strong coupling, the value of $\bar{r}$ is very small, allowing the energy to reach an extreme value. Therefore, we consider it a good approximation that $\bar{r}$ is small, i.e., $\lambda\gg \hbar/m_I\xi^2$. The second term of Eq.(\ref{IEe}) is much smaller compared to the first term due to
 \begin{equation}
   {\rm e}^{t^2}{\rm erfc}\left(t\right)\rightarrow 1, \ \ t\rightarrow 0.
 \end{equation}
Hence, we can simplify the integral result as follows:
\begin{equation}
 I_{\tilde{E}}\sim {\sqrt{2\pi}}/{2\bar{r}}.\label{fr}
 \end{equation}
 We notice that Eq.(\ref{fr}) does not depend on $\tilde{a}_{\pm}$. This demonstrates that, unlike in the weak coupling regime, phase separation has a negligible effect on the properties of polarons in the strong coupling regime. This is because, under strong coupling, the polaron forms a strongly self-trapping state that phase separation cannot disrupt. It is worth emphasizing that, the Fr\"{o}hlich model can only deal with strongly bound states resulting from the strong coupling of a single Bogoliubov excitation and an impurity. It has been suggested in some literature that the strong coupling of impurities and multiple Bogoliubov excitation lead to several times deeper bound states~\cite{Christianen2024}. Therefore, the model and method used in this paper can still capture the conclusion.

 In experiments, the scattering length between the impurity and the boson can be tuned up to resonance using Feshbach resonances. The results indicate that the effect of phase separation is negligible in the observation of strongly coupled Bose polarons. We expect that future experiments observe the Pekar bound state not only in one-component Bose gases, but also in two-component Bose gases under conditions of phase separation.

\section{Conclusion}\label{con}

In summary, We have analyzed the energies of polarons in single-component BEC and two-component BEC using the LLPH variational method, which allows us to obtain both weakly coupled and strongly coupled results. Interestingly, we find that phase separation has a significant effect on the properties of Bose polarons at weak coupling, but has little effect at strong coupling due to the formation of a strongly self-trapped state.

However, the LLPH approach has its limitations, such as its inability to account for intermediate coupling. Therefore, a complete phase diagram for all coupling strengths cannot be obtained using this method. Nevertheless, the advantage of this approach is that it combines results for weak and strong coupling to some extent, and it can also provide results for excited states~\cite{Chatterjee2019}.

\section{Acknowledgment}
The author would like to express gratitude for the helpful discussions with Prof. Zhan-Chun Tu and acknowledges the financial support received from the National Natural Science Foundation of China (Grant No. 11975050).

\end{CJK*}
\end{document}